\documentclass[prb,aps,amssymb,nofootinbib,twocolumn,showpacs]{revtex4}
\usepackage{amsmath}
\usepackage{amssymb}
\usepackage{amsthm}
\usepackage{amsfonts}
\usepackage{algorithmic}
\usepackage{enumerate}
\usepackage{latexsym}
\usepackage[dvips]{graphicx}

\newcommand{\beq}{\begin{equation}}
\newcommand{\eneq}{\end{equation}}
\newcommand{\beqs}{\begin{equation*}}
\newcommand{\eneqs}{\end{equation*}}
\newcommand{\vacket}{\left| \mbox{vac} \right>}
\newcommand{\vacbra}{\left< \mbox{vac} \right|}
\newcommand{\physket}{\left| \mbox{phys} \right>}

\newcommand{\half}{\frac{1}{2}}
\newcommand{\fdag}{f^{\dag}}
\newcommand{\cdagsup}{c^{\dag}_{\uparrow}}
\newcommand{\cdagsdown}{c^{\dag}_{\downarrow}}
\newcommand{\psidag}{\psi^{\dag}}
\newcommand{\su}{\uparrow}
\newcommand{\sd}{\downarrow}
\newcommand{\bi}{\mathbf{i}}

\newcommand{\anot}{a_{0}}
\newcommand{\xhat}{\hat{x}}
\newcommand{\yhat}{\hat{y}}

\input{epsf}

\begin{document}

\tolerance 10000

\title{Inhomogeneous states and nodal fermions in the $SU(2)$ gauge theory}

\author {B.W.A. Leurs, K.E. Luna and J. Zaanen}

\affiliation{ Instituut Lorentz for Theoretical Physics, Leiden University, Leiden, The
Netherlands}

\begin{abstract}
\begin{center}
We discuss the issue of phase separation in the $SU(2)$ slave boson theory of Wen and Lee of
the doped Mott insulator \cite{wenlee}. It is shown that the constraint structure of the
theory leads to the interpretation of the holons to have hard-core interactions, which is
demonstrated further by studying the empty limit (no electrons). Surprisingly, with hard-core
interactions even the empty limit is described well by the slave-boson theory, both as an
energy density and with the regard to dynamical properties. The consequences are investigated
in the overdoped superconducting regime, where both phase separation and a $d+s$ structure of
the order parameter is obtained. This $s$-wave component is already imminent in the
description of the hole in the slave boson theory. The interacting nature of the holons also
lead to sound modes in the single-electron propagator. The novel idea of the isospin spiral is
introduced, based on the projective symmetry principles of Wen. This isospin spiral explains
the coexistence of superconductivity and the Mott insulating state, being the consequence of
phase separation. Secondly, it might be able to explain why nodal fermions survive in the
presence of charge inhomogeneities.
\end{center}
\end{abstract}

\date{\today}

\pacs{}

\maketitle

\section{Introduction} \label{sect:intro}

A long standing question in the physics of high-$T_c$ superconductivity is how nodal fermions
can coexist with stripes, as experiments seem to point out \cite{seamus}. There is no simple
theoretical explanation on the market as to why charge inhomogeneities do not interact
strongly with the low-lying degrees of freedom. A possible explanation is that in the
high-$T_c$'s, electrons are splintered into spinons and holons. Being different degrees of
freedom, it is possible that charge inhomogeneities do not communicate with the fermionic
low-lying excitations. Experimental support for such fractionalisation is for example given by
photo-emission \cite{orgad2001}.

This idea of fractionalisation is incorporated in the slave boson theories, leading to spin
liquid states \cite{bza,chiralsl}. These spin liquids are states ``nearby" the parent Mott
insulating states, and carry nodal excitations \cite{affleck2}. The hope is that these spin
liquid states are candidates to describe the evolution from the Mott insulator to the $d$-wave
superconductor, in which quasiparticles with $d$-wave dispersion are extensively proven
experimentally \cite{arpesreviewshen}.

These slave theories are plagued by one problem, however: they appear not to be able to
describe inhomogeneous states, like stripes, let alone that they can explain coexistence of
nodal fermions with stripes. The main point of this article is that the $SU(2)$ slave boson
theory as developed by Lee and Wen \cite{wenlee, lnw,lnwng}, is able to capture both aspects.
This conclusion is based both on the projective symmetry ideas of X.-G. Wen \cite{wenpsg}, and
on an improvement of the original theory by Wen and Lee.

The first idea is at the root of our proposal of the \emph{isospin spiral}. Wen showed that
for zero doping, the mean field states describing a staggered flux phase (SFP) \cite{affleck1}
and the $d$-wave superconductor (dSC), are physically the same, i.e., they are gauge
equivalent. Only for non-zero doping, this equivalence is broken, such that a $d$-wave state
is favoured, mimicking the instability of the SFP against the dSC \cite{kotliar}. Our idea is
now that for low doping, an inhomogeneous state is favourable above a homogeneous dSC. This
state connects an SFP with no doping smoothly with a charged superconductor. This leads to a
picture in which isolating SFP states coexist with dSC states on stripes. The protection of
the nodal fermions lies in the fact that for zero doping, both the SFP and dSC carries nodal
excitations.

For these inhomogeneous states to exist, it needs to be proven that the $SU(2)$ slave boson
theory supports phase separation, which is not proven so far. We show that this phase
separation does occur, which is connected to our technical improvement of the original Wen-Lee
theory. It is demonstrated that due to the constraint structure of the theory, the charge
carrying holons need to have a hard-core interaction, which accounts for phase separation. The
predicted compressibility and critical doping are in accord with chemical potential shift
measurements \cite{fujimori}. These phase separation tendencies of the $SU(2)$ gauge theory
puts the door ajar for more intricate phenomena, like the stripe order \cite{zaanen,
kivelsonPS, white} that has been observed in experiments \cite{tranquada95, tranquada97,
yamada98, mook98, balatsky99}.

In turn, this constraint structure turns out to be responsible for an unexpected surprise: the
superconducting order parameter \emph{needs} to be $d+s$-wave, instead of $d$. This interferes
in an interesting way with the empirical developments in high-$T_c$ superconductivity. There
is strong experimental evidence from $c$-axis tunneling \cite{klemm2005}, \cite{klemm2006} and
Raman scattering \cite{tajima,nemetschek} that in Bi2212 there is an $s$-wave component in the
gap, which is in largeness comparable \cite{klemm2006} to our prediction, and grows with
doping \cite{tajima}, also in accord with our prediction. Further, $\pi$-phase shift
experiments for YBCO point out that the $s$-wave component therein cannot be fully explained
by the orthorombicity of the crystal \cite{hilgenkamp, smilde}. As far as we are aware,
$SU(2)$ gauge theory, in our formulation, is the only  theory explaining these results in at
least an elegant way, and is demonstrated to be rooted in the way $SU(2)$ gauge theories
describe doping of holes.

It appears that the above experimental findings are largely ignored because all existing
mechanism theories predict either a $d$-wave or an $s$-wave, and the $SU(2)$ gauge theory is
stand-only with regard to its insistence on a $d+s$-symmetry. In the narrow context of
slave-like theories, Ogata and coworkers excluded $d+s$ in the related context of Gutzwiller
projected wave function Ansatzes \cite{ogata99,ogata2000}.

The organisation of this paper is as follows. In Section \ref{sect:su2form}, we review the
$SU(2)$ gauge theory formulation of Wen and Lee, and mention the projective symmetry ideas of
Wen. In Section \ref{sect:empty}, we take a somewhat warped perspective on doping: to fully
understand the effects of doping, we consider the empty limit, i.e., the doping $x=1$. As
academic as it might seem, this exercise is very instructive as to the fact that the slave
holons should have hard cores, and that the $s$-wave order parameter is induced by doping. It
turns out that the $s$-wave component is equal to half times the density of dopant holons. We
also show that the mean field theory is able to get both the energy and the single-electron
propagator right. Furthermore, the mean field theory in the empty limit gives the inspiration
for the mean field wave function to be exploited in our analysis for intermediate dopings in
the following section.

In that Section \ref{sect:psds}, we derive a mean field free energy functional for non-zero
doping in the grand canonical ensemble, to be able to account for phase separation. The mean
field phase diagram is calculated. The phase separation properties of the hard-core holon
$SU(2)$ slave theory is further quantified by comparison of the compressibility with
experiments on the chemical potential shift \cite{fujimori}. The agreement turns out to be
very well. Also the node shift, caused by the $s$-wave admixture, is determined. The hard-core
nature of the bosons leads to interesting and experimental falsifiable properties. Namely, the
hole condensate leads to phonon modes in the incoherent part.

In the last Section \ref{sect:isospin} the idea of the isospin spiral is introduced and
quantified. We show that the mean field energy of the inhomogeneous state is only a bit above
the energy of the homogeneous state. We argue that this should be viewed as an artefact of the
$t-J$-model, and not of the isospin spiral state. We also show that the nodal fermions are not
that much affected by the charge inhomogeneities as expected. In effect, the gap is very
small, leading to only very small Umklapp scattering at the Fermi pockets.

These interesting results form a promising motivation to study $SU(2)$ gauge theory in more
realistic models for the cuprates, the more so since it seems to be the only theory on the
market predicting a $d+s$-form of the superconducting order parameter. In fact, $SU(2)$ gauge
theory forms the bridge between, on the one hand, nodal fermions and spin liquid states, and
on the other hand, striped superconductors.

\section{$SU(2)$-slave boson formulation of the $t-J$-model} \label{sect:su2form}

As is widely accepted, the problem of high-temperature superconductivity, is the problem of
doping a Mott insulator. The parent compounds of high-$T_c$ are insulating, due to a large
Coulomb repulsion. By removing electrons however , the charges  get mobile, and it is believed
that this physics is at the origin of the superconductivity. This forms the motivation to
include the hopping of projected electrons $\tilde{c}_{i\sigma}$ in the original Heisenberg
Hamiltonian:
\begin{equation}
\label{tJham} H_{t-J} = \sum_{< ij >} J \left( \mathbf{S}_{i}\cdot \mathbf{S}_{j} -
\frac{1}{4}n_{i}n_{j} \right) - \sum_{ij} t_{ij}
\left(\tilde{c}^{\dag}_{i\sigma}\tilde{c}_{j\sigma} + h.c. \right),
\end{equation}
where the hoppings $t_{ij}$ are the wave function overlaps of electrons at sites $i$ and $j$.
Without loss of generality, we take $t_{ij}=t$ for nearest neighbour sites, and $t_{ij}=0$
otherwise.

The idea of the $SU(2)$ slave boson formulation, due to Wen and Lee\cite{wenlee,lnw}, is
rooted in the observation by Affleck and Marston \cite{affleck1} that the spin system at
half-filling in a fermionic 'spinon' representation is characterised by both the usual 'stay
at home' $U(1)$ gauge symmetry, and a local conjugation symmetry meaning that one can describe
the spin system equally well in terms of spinon particles and antiparticles. This idea is
encoded in introducing the $SU(2)$ doublet composed from the spinon operators $f_{i\su}$ and
$f_{i\sd}$,
\begin{equation} \label{psi}
\psi_i =\left (
      \begin{matrix}
        f_{i\su} \\
      \fdag_{i\sd}
      \end{matrix}
      \right).
\end{equation}

In this language, the spin operators are given by
\begin{eqnarray}
\mathbf{S}^+_i &=& \half( \psidag_{1i}\psidag_{2i} - \psidag_{2i}\psidag_{1i} ) \\
\mathbf{S}^z_i &=& \half( \psidag_i \psi_i - 1 ).
\end{eqnarray}

Importantly, the Hilbert space of the $t-J$ Hamiltonian is formed from
 three states only: a spin up electron, $\cdagsup \vacket $, a spin down electron
$\cdagsdown \vacket $ and a vacancy $\vacket $. Consequently, the electron operators in
Eq.(\ref{tJham}) are projected electron operators $\tilde{c}_{i\alpha} =
c_{i\alpha}(1-n_{i\bar{\alpha}})$, where $\bar{\alpha}$ denotes a spin opposite from $\alpha$.
The tilde is dropped from now on and  the projection is kept implicit. Bear in mind that one
should take care that all physics takes place in this projected Hilbert space!

Let us now describe these electrons in the $SU(2)$ gauge theory. Since in everyday life only
the physical electrons are encountered, the electrons should be $SU(2)$ singlets  To construct
these singlets, introduce the $SU(2)$ doublet describing holons:
\begin{equation} \label{holons}
h_i =\left (
      \begin{matrix}
        b_{1i} \\
        b_{2i}
      \end{matrix}
      \right).
\end{equation}
Then the appropriate $SU(2)$ singlet describing the projected electron is
\begin{eqnarray} \label{cparam}
c_{\uparrow i} &=& \frac{1}{\sqrt{2}} h^{\dag}_i \psi_i =
 \frac{1}{\sqrt{2}}( b^{\dag}_{1i}f_{\uparrow i} + b^{\dag}_{2i}f^{\dag}_{\downarrow i} ) , \\
 c_{\downarrow i} &=& \frac{1}{\sqrt{2}} h^{\dag}_i \overline{\psi}_i =
 \frac{1}{\sqrt{2}}( b^{\dag}_{1i}f_{\downarrow i} - b^{\dag}_{2i}f^{\dag}_{\uparrow i} ).
\end{eqnarray}

The equations (\ref{cparam}) as they stand, however, are not operator equalities, since the
Hilbert space of the $SU(2)$ theory is larger. To arrive at the correct physics, we have to
impose constraint to make the mapping to the states of the original $t-J$-Hamiltonian exact,
so that the Hilbert spaces are equal. This is achieved as follows. Since the electrons should
be $SU(2)$ singlets, we should require that the physical states $\physket$ of the $SU(2)$
slave boson model obey \cite{wenlee}
\begin{equation}\label{exactconstr}
\left (\psidag_i \tau^l \psi_i + h^{\dag}_i\tau^l h_i \right)\physket = 0.
\end{equation}

There are precisely three states satisfying Eq. (\ref{exactconstr}). The first two states
satisfying those, are $\fdag_{\su i}\vacket$ and $\fdag_{\sd i}\vacket$, corresponding the
projected up and down electron in the $t-J$ model. The description of the hole leads
inevitably to the introduction of the holon doublet Eq.(\ref{holons}). This is seen by the
fact that the empty state $\vacket$ is conversed into a spinon doublet by the projection
operator in Eq. (\ref{exactconstr}), and vice versa. Henceforth, when an empty site is
accompanied with a $b_1$ boson,
 and a doubly occupied site with a $b_2$ boson, the hole in the $t-J$ model is
 represented by
 \begin{equation}\label{holereps}
\left| 0 \right>_i = \frac{1}{\sqrt{2}}\left(b^{\dag}_{1i} + b^{\dag}_{2i}\fdag_{\sd
i}\fdag_{\su i} \right)\vacket_i.
 \end{equation}
It is easy to check that Eq. (\ref{holereps}) \textit{does} satisfy all the constraints Eq.
(\ref{exactconstr}). Observe that we needed all the three constraints to arrive at this
expression. In the original Wen and Lee formulation, the consequences of this fact have not
been taken at face value. It will be demonstrated that this is not justified for higher
dopings.

In the expression for the hole, $SU(2)$ gauge theory already reveals some of its powers: it
captures the fact associated with the particle-hole symmetry intrinsic to spin that the
singletness of pure vacuum should be treated on precisely the same footing as the spin
singletness of either the empty or doubly occupied spinon configuration.

In this way, it is shown how one can make a mapping from the slave boson operator states to
the Hilbert space of the $t-J$ model, by including exact constraints Eq.(\ref{exactconstr}).
Solving for exact constraints is however extremely difficult.

In order to make progress, we are going to put forward a mean field theory, and treat the
constraints on a mean field level.

one may assume the existence of the fermionic vacuum expectation values
 \begin{eqnarray}
\chi_{ij}   &=& \langle \fdag_{i\su}f_{j\su} + \fdag_{i\sd}f_{j\sd} \rangle \label{chi} \\
\Delta_{ij} &=& \langle f_{i\su}f_{j\sd} - f_{i\sd}f_{j\su} \rangle. \label{Delta}
 \end{eqnarray}
 The first one is a hopping amplitude, inspired by the staggered flux spin liquid states proposed by
Affleck and Marston \cite{affleck1,affleck2}. The order parameter $\Delta$ is going to play
the role of a superconducting amplitude with $d$-wave symmetry.

Let us now assume that the expectation values Eq. (\ref{chi}) and Eq. (\ref{Delta}) exist, and
let us presuppose deconfinement, by neglecting fluctuations in the gauge fields $a^{l}_{0i}$.
This is equivalent to replacing of the exact constraints Eq. (\ref{exactconstr}) by the mean
field constraints
\begin{equation}
\label{mfconstr} \langle \psi^{\dag}_i \tau^{l} \psi_i \rangle = 0.
\end{equation}

Let us first obtain a manifestly invariant $SU(2)$ gauge invariant mean field theory, by
grouping the mean fields as follows:
\begin{equation} \label{uij}
U_{ij} = \left (
      \begin{matrix}
   -\chi_{ij}^*   & \Delta_{ij} \\
    \Delta_{ij}^* & \chi_{ij}
      \end{matrix}
      \right).
\end{equation}

To actually calculate matters, we have to derive the slave boson version of the Hamiltonian
Eq.(\ref{tJham}), with the decomposition Eq.(\ref{cparam}). To decouple terms quartic in the
spinons, we use the spin liquid Ansatzes Eqns.(\ref{chi}) and (\ref{Delta}). In order to
impose the three constraints Eq.(\ref{dopedmfconstr}), we need to incorporate the Lagrange
multipliers $a^l_{0i}$ into the mean field Hamiltonian. Bearing these remarks in mind, it is a
straightforward exercise to derive the mean field Hamiltonian
\begin{eqnarray}
H_{mf} &=& -\mu \sum_{i} h^{\dag}_{i}h_{i} - \sum_{i} a_{0i}^{l}
\left(\frac{1}{2}\psi^{\dag}_{\alpha
i}\tau^{l}\psi_{\alpha i} + h^{\dag}_{i}\tau^{l}h_{i} \right) \nonumber \\
& & + \ \ \sum_{< ij >} \frac{3J}{8} \left( |\chi_{ij}|^2 + |\Delta_{ij}|^2
              +\psi^{\dag}_{i}U_{ij}\psi_{j}  + h.c. \right) \nonumber \\
& & + \ \ \sum_{<ij>}  t\left( h^{\dag}_{i}U_{ij}h_{j} + h.c. \right), \label{mftheory}
\end{eqnarray}
where $U_{ij}$ was already defined in Eq.(\ref{uij}).

Of course, the Hamiltonian Eq.(\ref{mftheory}) is manifestly $SU(2)$ invariant under the
transformation
\begin{eqnarray} \label{dopedtrafo}
& &\psi_i \rightarrow g_i\psi_i, \mbox{\ \ } h_i \rightarrow g_i h_i, \mbox{\ \ }U_{ij}
\rightarrow \tilde{U}_{ij}=g_i U_{ij} g^{\dagger}_j, \nonumber \\
& & \mbox{\ \ } a^l_{i0}\tau^l \rightarrow \tilde{a}^l_{0i} \tau^l = g_i a^l_{i0}\tau^l
g^{\dagger}_j.
\end{eqnarray}

Inspired by the spin liquid ideas of many people \cite{pwa,lee99, senthil1},
 an idea having some experimental support \cite{orgad2001} , we introduce
 three mean field states, namely the staggered flux phase \cite{affleck1, affleck2}, the
$d$-density wave state \cite{chetanddw, sudiphiddenorder} and the $d$-wave superconductor. We
recall their descriptions here.

The $d$-wave superconductor is in the projective symmetry group represented by
\begin{eqnarray}
 \mbox{dSC} \nonumber & & \\
U_{\bi, \bi + \xhat} &=& -\chi\tau^3 + \Delta\tau^1, \nonumber \\
U_{\bi, \bi + \yhat} &=& -\chi\tau^3 - \Delta\tau^1, \nonumber \\
\anot^3 &=& 0, \ \  \anot^{1,2} \not= 0, \nonumber \\
<b_1> &=& <b_2> \not= 0.
\end{eqnarray}

The dispersion for the fermions is (at least for homogeneous $a_{0i}$) readily calculated to
be
\begin{eqnarray}\label{Ek}
E_k = \sqrt{(\chi_k - \anot^3)^2 + (\Delta_k - a^1_0)^2 }, \nonumber \\
\chi_k = -\frac{3J}{4}(\cos k_x + \cos k_y)\chi, \nonumber \\
\Delta_k = -\frac{3J}{4}(\cos k_x - \cos k_y)\Delta.
\end{eqnarray}
One should notice that the spinons have gapless Dirac dispersions at the points
 $(k_x,k_y)=(\pm \half \pi, \pm \half\pi )$, i.e., at those points we have nodal fermions
 consistent with experiments
  \cite{ding95_1, ding95_2}.

 The following two phases also support these Dirac quasiparticles. The first one is the
staggered flux phase,
\begin{eqnarray}
 \mbox{SFP}& &\nonumber \\
U_{\bi, \bi + \xhat} &=& -\chi\tau^3 - i(-)^I \Delta, \nonumber \\
U_{\bi, \bi + \yhat} &=& -\chi\tau^3 + i(-)^I \Delta, \nonumber \\
\anot^{1,2,3} &=& 0, \nonumber \\
<b_1> &=& <b_2> = 0.
\end{eqnarray}

The third phase has Fermi pockets, with nodes radially shifted from $(\half \pi, \half \pi)$:
\begin{eqnarray}
\mbox{dSC with pockets} & & \nonumber \\
U_{\bi, \bi + \xhat} &=& -\chi\tau^3 - i(-)^I \Delta, \nonumber \\
U_{\bi, \bi + \yhat} &=& -\chi\tau^3 + i(-)^I \Delta,\nonumber \\
\anot^{1,2} &=& 0, \anot^3 \not=0 \nonumber \\
<b_1> &\not=0&, <b_2> = 0.
\end{eqnarray}
We remind the reader that the $dSC$ state is referred to that way, since in the Hamiltonian
Eq. (\ref{mftheory}) that particular $U_{ij}$ couples $f_{\su i}$ with $f_{\sd i}$.

The SFP is a spin liquid, describing spinons hopping around the plaquettes of the square
copper-oxide lattice. It breaks translation symmetry, since the hopping fluxes
\begin{equation}\label{plaqflux}
\Phi_{hop} = \frac{\pi}{4} \sum_{\mbox{\tiny plaquette}} \mbox{Arg}( U^{11}_{ij}) = \pm
\frac{\Delta}{\chi}\pi
\end{equation}
show a bipartite staggered pattern. By a Fourier transformation to momentum space, the
dispersion is readily calculated. It turns out to be identical to the dSC-dispersion
\begin{eqnarray}
E_k = \sqrt{(\chi_k - \anot^3)^2 + (\Delta_k - a^1_0)^2 }, \nonumber \\
\end{eqnarray}
The three mean field phases above describe nodal fermions, but at different points in
$k$-space.

For zero doping, however, all the three Ansatzes become the same, supporting Dirac
quasiparticles at $(\half\pi, \half\pi)$, since the Lagrange multipliers and boson densities
vanish.  This is surprising, since the SFP breaks translation symmetry, whereas the dSC does
not. In the framework of classical Landau-Ginzburg-Wilson theory, this is impossible: in
general, different symmetry broken states give rise to different excitations. What is going on
here? In fact, the two above states are two sides of the same coin. This is seen after
applying the following site-dependent transformation,
\begin{equation}\label{gispecial}
g_i = \exp\left(-i\frac{\pi}{4}(-)^I \tau^1\right)
\end{equation}
by which the SFP is mapped to the dSC. (The quantity $I$ is defined as $I=i_x+i_y$.) Put
differently, the translational symmetry breaking of the SFP is just a gauge artefact.

Due to the idea of Wen \cite{wenpsg}, this is an expression of the fact that for zero doping,
these states are members of the same projective symmetry group ($PSG$). This means that
different mean field states describing the same physics, are connected to each other by gauge
transformations. Wen then declares them to have the same \emph{quantum order}, a novel concept
going beyond the standard Ginzburg-Landau-Wilson paradigm.

In fact, all states connected to the dSC by a gauge transformation
\begin{equation}
g_i = \exp\left(-i\theta_i \tau^1\right)
\end{equation}
are equivalent. This can be pictured nicely by the concept of the isospin sphere. An
$SU(2)$-gauge group element $g_\bi$ can be written as follows:
\begin{equation} \label{gi}
g_\bi =
 \left (
      \begin{matrix}
     z_{\bi 1} & -z^{*}_{\bi 2}  \\
     z_{\bi 2} &  z^{*}_{\bi 1}
      \end{matrix}
 \right)
\end{equation}
where the complex numbers $z\bi$ are parametrised by three angles, viz.,
\begin{equation} \label{zi}
z_{\bi 1} = e^{i\alpha_\bi}e^{-i\frac{\phi_\bi}{2}}\cos\frac{\theta_\bi}{2}, z_{\bi 2} =
e^{i\alpha_\bi}e^{ i\frac{\phi_\bi}{2}}\sin\frac{\theta_\bi}{2}.
\end{equation}
The $z$'s are grouped in the vector $z_\bi = (z_{\bi 1}, z_{\bi 2})$.

The isospin vector $\mathbf{I}_\bi$ turns out to be a useful definition:
\begin{equation} \label{I}
\mathbf{I}_\bi = z_\bi^{\dag}\mathbf{\tau}z_\bi = (\cos{\phi_\bi}\sin{\theta_\bi},
\sin{\phi_\bi}\sin{\theta_\bi}, \cos{\theta_\bi})
\end{equation}

\begin{figure}
\label{isospin}
\centering \rotatebox{0}{
\resizebox{6.3cm}{!}{%
\includegraphics*{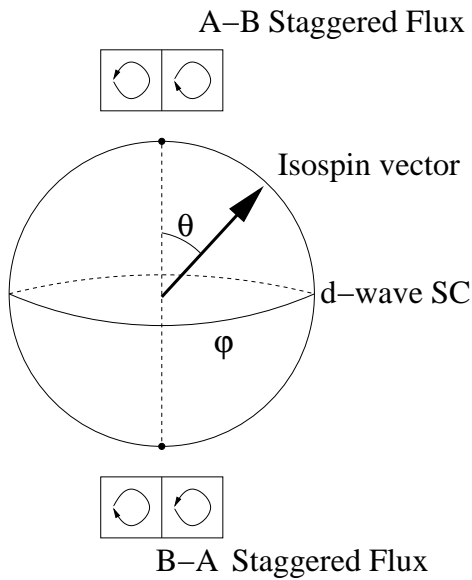}}}
\caption{\small{The isospin sphere. States on the north and south pole are staggered flux
phases, whereas states on the equator are dSC. In between, we have a DDW state. }}
\end{figure}

The angle $\theta$ can then be interpreted as the latitude on the isospin sphere, whereas the
angle $\phi$ is the longitude, cf. Figure \ref{isospin}. The north and south pole of the
sphere correspond to a staggered flux phase, with $A-B$ and $B-A$ staggering respectively,
while the equator corresponds to the $d$-wave superconductor. For half filling, the rotations
on the isospin sphere correspond to pure gauge transformations, meaning that spinon flux
phases and d-wave superconductors are gauge-equivalent.

The way the $SU(2)$ mean field theory is set up, is as follows. The spinons and holons are
considered to be separate systems. As long as one rotates the spinons and holons together,
$SU(2)$ gauge symmetry gives the same mean field properties. If one fixes a gauge for the
spinons, and then starts to rotate the holons independently, the mean field results for the
energy will be different. The strategy we choose is to fix the spinon gauge at the dSC mean
field state, whereas the holons will be rotated by the group element $g_i$. It was shown in
Eq.(\ref{gi}) that  $g_i$ can be decomposed into ``Euler angles'' $\alpha_i, \theta_i$ and
$\phi_i$. These can be encoded in the useful concept of isospin, as defined in (\ref{I}). We
pictured this concept in the isospin sphere, cf. Fig. \ref{isospin}. Only the angle $\theta_i$
will turn out to be physical, whereas the other ones are gauge. For equator states, we have
$\theta=\half\pi$, corresponding to $<b_1>=<b_2>$, whereas for $\theta=0$, $<b_1>\not=0$ and
$<b_2>=0$. The latter means that the symmetry between empty sites and doubly occupied sites is
broken.

The concept of isospin will turn out to be important for the last section, since it can encode
for inhomogeneous states as $\theta_i$ might vary with lattice site $i$. But before turning to
that topic, we need to cross some terrain.

\section{Lessons from the empty limit}\label{sect:empty}

In this section, we investigate the consequences of the peculiar form of the expression for
the hole in the $SU(2)$ formulation. We will demonstrate in the first subsection that the
spinon pair involved will give rise to an $s$ wave component in the spinon pairing parameter.
Furthermore, we will show that the constraint equations lead to requirement of the bosons to
have a hard core. This property will also take care of the mean field energy to be correct in
the empty limit. In the last subsection it is demonstrated that the proposed mean field wave
function even renders the single-electron propagator correctly.

Let us first discuss the hard-core nature of the holons. The exact wave function describing
the empty limit is simple,
\begin{equation}\label{emptyexact}
|0\rangle = \prod_i \left( b^{\dagger}_{i1 } +  b^{\dagger}_{i2} f^{\dagger}_{\sd i}
f^{\dagger}_{\su i } \right) \vacket.
\end{equation}
Deconfinement or spin-charge separation implies that the system loses its knowledge about the
three particle correlation $b^{\dagger}_{i2} f^{\dagger}_{\sd i} f^{\dagger}_{\su i} $  and
the best one can do is to look for a holon-spinon product wave function. The best choice is
obviously
\begin{equation} \label{emptyMFstate}
 | 0 \rangle_{MF} = \prod_i \left[ \frac{1}{2} \left( b^{\dagger}_{i1 } + b^{\dagger}_{i2}
\right) \left( 1 + f^{\dagger}_{\sd i} f^{\dagger}_{\su i} \right) \right] \vacket .
\end{equation}

This wave function still has to satisfy the mean-field version of the constraint Eq.
(\ref{exactconstr}),
\begin{equation}\label{dopedmfconstr}
\left< \psidag_i \tau^l \psi_i + h^{\dag}_i\tau^l h_i \right>  = 0,
\end{equation}
where the brackets in this case stand for the expectation value relative to the state $| 0
\rangle_{MF}$. This brings us to the main point: the constraints are only satisfied when  the
bosons have hard cores. Indeed, the mean field wave function Eq. (\ref{emptyMFstate}) obeys
\begin{equation}
\left< 0 \right| \psidag_i \tau^3 \psi_i \left| 0 \right>_{MF} +\left< 0 \right| h_i \tau^3
h_i \left| 0 \right>_{MF} = 0 + 0 = 0.
\end{equation}
If we were to take soft-core bosons, the state $\left(  (b^{\dag}_{1i})^2 + b^{\dag}_{2i}
\right)\vacket $ would be possible. It does not satisfy the constraints, however, since then
\begin{equation}
\left< 0 \right| \psidag_i \tau^3 \psi_i \left| 0 \right>_{MF} +\left< 0 \right| h_i \tau^3
h_i \left| 0 \right>_{MF} = 0 + 1 \not= 0.
\end{equation}
So, if we are to take the Hilbert space constraints seriously, we need to accept that the
holons have an infinite hard core. This is consistent with the fact that the holons carry
electric charge. Since the Coulomb repulsion is taken infinite to arrive at the $t-J$ model in
the first place, this means that there can be at most one $b$-boson per site.

The reason to stress this hard-core nature of the bosons, is that in the original formulation
\cite{wenlee}, the bosons were taken to be non-interacting. The empty limit exercise, being
transparent in this regard, shows that this is inconsistent for appreciable dopings.  This is
also illustrated by the fact that the hard-core bosons give the correct energy in the empty
limit. Indeed, let us first show that with the original projected electrons, the exact energy
of the $t-J$ Hamiltonian Eq. (\ref{tJham}) is zero in this limit. The empty state is described
in the projected electron formulation simply by $\vacket$, implying vanishing $n_i$ on this
state. Since the spin operator $\mathbf{S}^l_i$ is given by
$c^{\dag}_{i\alpha}\half\tau^l_{\alpha\beta}c_{i\beta}$, the energy of the  Heisenberg part
vanishes. Since there are no electrons around, the hopping part vanishes as well, and we
conclude that the total energy vanishes.

Let us now demonstrate that the Ansatz Eq.(\ref{emptyMFstate}) yields the same result.
Firstly, the Heisenberg term vanishes. The only component of spin that could contribute is the
$l=3$ component, since the others vanish when acting on $\left( 1 + f^{\dagger}_{\sd i}
f^{\dag}_{\su i} \right) \vacket $ . However, since $\mathbf{S}^3_i$ is the number of up spins
minus the number of down spins, it vanishes as well. Furthermore, since the number operator
$n_i$ in the slave boson representation reads $c^{\dag}_{i\alpha}c_{i\alpha}= \half
\psi^{\dag}_i h_i h^{\dag}_i \psi_i$, it also gives zero contribution, because of the hard
core condition. Similarly, the hoppings vanish for the same reason.

This would not be the case if the bosons were assumed to have no hard core, since for weakly
interacting bosons there is no restriction on the hoppings. This is inconsistent with the
$t-J$ model, as hopping is only allowed between occupied and empty sites. In conclusion, the
hard core condition is a sufficient condition for the empty state to have the correct energy
in the empty limit.

So far, we have derived an exact expression for the hole creation operator in the $SU(2)$
gauge theory, taking seriously all three constraints. We considered the empty limit next,
since it is easy to construct the mean field theory in this case. Treating the constraints
correctly, we arrived at the conclusion that the bosons should be treated as having a hard
core. In the next section, we will show how our mean field Ansatz Eq.(\ref{emptyMFstate})
generalises to  mean field wave functions describing intermediate dopings.

\subsection{Doping induces an $s$-wave order parameter}\label{subsect:swave}

In the previous section, we showed that a correct description of the empty limit requires that
the bosons should be treated as hard-core.  The empty limit considerations leading us to that
conclusion, turns out to be very useful to find out the structure of the mean field wave
function at intermediate dopings. In fact,
 hard-core bosons are
just like XY spins, and the straightforward generalisation of Eq.(\ref{emptyMFstate}) becomes
obvious,
\begin{equation}
| \Psi_0 \rangle_{holons} = \prod_i \left( \alpha_i + \beta_{i} (u_i   b^{\dagger}_{i1 } + v_i
b^{\dagger}_{i2 } ) \right) \vacket \label{holonMFstate},
\end{equation}
where the complex numbers $\alpha$ and $\beta$  obey the normalisation condition
$|\alpha|^2+|\beta|^2=1$.

To already harvest some results from our considerations, let us consider the saddle point
Lagrange multiplier equations
\begin{equation}
\frac{\partial}{\partial a^{l}_{0}} \left< H_{mf} \right> = 0, \mbox{\ \ \ } l = 1,2,3.
\end{equation}
These are precisely the mean field constraint equations Eq.(\ref{dopedmfconstr}):
\begin{eqnarray}
\label{mfconstraints} \left< f^{\dag}_{\uparrow i} f^{\dag}_{\downarrow i} + f_{\downarrow i}
f_{\uparrow i} \right> &=& \left< b^{\dag}_{1i}b_{2i} + b^{\dag}_{2i}b_{1i}\right>
\\
-i\left< f^{\dag}_{\uparrow i} f^{\dag}_{\downarrow i} - f_{\downarrow i} f_{\uparrow i}
\right> &=& -i\left< b^{\dag}_{1i}b_{2i} - b^{\dag}_{2i}b_{1i} \right>
\\
\left< f^{\dag}_{\alpha i} f_{\alpha i} -1 \right> &=& \left<b^{\dag}_{2i}b_{2i} -
b^{\dag}_{1i}b_{1i}\right>.
\end{eqnarray}
These constraint equations already convey an important message. The third equation tells us
something about the deviation from half-filling, which was an important point already made by
Lee, Wen and Nagaosa \cite{lnw}. As soon as the average fermion occupation number deviates
from unity, i.e., deviates from half filling, there is a difference between $b_1$ bosons and
$b_2$ bosons. In plain physics language: as soon as Fermi pockets form, the difference between
empty sites and spin pair singlets becomes physical.

The first two equations acquire a novel interpretation. For non-zero dopings, the holon
expectation values are non-zero as well. However, looking at the left-hand side of the
equations, one needs to conclude that a superfluid order parameter appears with an $s$-wave
structure. In other words, taking seriously all constraint equations, and having convinced
oneself that doping must be described by a superposition of both empty and doubly occupied
sites, one has to face an extra order parameter with a superconducting $s$-wave symmetry.
Rephrased in physical language: within the framework of $SU(2)$ theory, doping induces
$s$-wave pairing.

One could argue that there are some left-over degrees of freedom, so that one could gauge away
the $s$-wave component. This is not the case, however. Let us exploit the isospin
representation, introduced in Section \ref{sect:su2form}. Using the isospin angles $\varphi$
and $\theta$, cf. Fig. \ref{isospin}, we parametrise the holon wave function
(\ref{holonMFstate})
 by  $u_i = \cos(\frac{\theta_i}{2})$ and $ v_i = \sin(\frac{\theta_i}{2})$. Further, choose
$\beta_i \rightarrow \beta_{i} e^{i\varphi_i}$ such that $\beta_{i}$ is real. This
parametrisation is instructive, since $\theta_i = \frac{\pi}{2}$ makes the expectation values
for $b_1$ with vacancies indistinguishable from $b_2$ with a spinon pair, reproducing the
particle-hole symmetric empty state Eq.(\ref{emptyMFstate}). Moreover, this corroborates the
point that the equator on the $SU(2)$-isospin sphere (i.e.,$\theta_i = \frac{\pi}{2}$ )
corresponds to the particle-hole symmetric d-wave superconductor. Calculating the expectation
values explicitly, the equations Eq.(\ref{dopedmfconstr}) become
\begin{eqnarray}
\left< f^{\dag}_{\uparrow i} f^{\dag}_{\downarrow i} + f_{\downarrow i} f_{\uparrow i} \right>
&=& |\beta_{0i}|^{2} \sin(\theta_i)\cos(\varphi_i) \label{constr1}
\\
\left< f^{\dag}_{\uparrow i} f^{\dag}_{\downarrow i} - f_{\downarrow i} f_{\uparrow i} \right>
&=& |\beta_{0i}|^{2} \sin(\theta_i)\sin(\varphi_i) \label{constr2}
\\
\left< f^{\dag}_{\alpha i} f_{\alpha i} -1 \right> &=& |\beta_{0i}|^{2} \cos(\theta_i).
\label{constr3}
\end{eqnarray}
On the one hand, this  illustrates once again that equator states are particle-hole symmetric,
cf. Eq.(\ref{constr3}). Secondly, and more importantly, we see that there is no way to gauge
away the $s$-wave component. In other words, as soon as there is a superconducting order
parameter $(\sin(\theta)\not= 0)$, there is an $s$-wave component linearly increasing with
doping $x$:
\begin{equation}
\Delta_s = \half x \sin(\theta).
\end{equation}

 The only freedom is to choose its phase to be real by choosing $\varphi_i=0$,
implying zero $a^2_{0i}$ \cite{wenpsg}. This is a first result of our empty-limit exercise,
which at first sight looks trivial. Conversely, the first equation tells us that we cannot
neglect the Lagrange multiplier $a^1_{01}$, which accounts for the $s$-wave admixture. This
has been ignored in the original formulations of the mean field theory \cite{lnw, wenlee,
lnwng}. This flaw leads to a disaster, as we will show in the next subsection.

\subsection{The empty limit in mean-field theory} \label{subsect:emptymf}

One could wonder if it is really wrong to leave out the first constraint. Probably one remains
very closely to the "correct" mean field state when ignoring it? This is not the case: it
leads to nonsensical results. The bright side is the ease by which the constraint $a^1_{0i}$
is incorporated. Let us fix the holon density $\left< h^{\dag}_i h_i \right> =1 $, and choose
the Hubbard-Stratonovich fields to be homogeneous, $\Delta_{ij} = \Delta, \chi_{ij}=\chi $.
Since the empty state makes no distinction between empty and doubly occupied sites, we have
$\theta_i =\theta = \half \pi $. Inserting these assumptions into Eq.(\ref{mftheory}), we
obtain an energy density functional $E_{mf}$ for the empty limit.

Let us first ignore $a^1_{01}$. Then the mean field equations are
\begin{eqnarray}
\frac{\partial E_{mf}}{\partial \chi} &=& 2  \chi  = \sum_k \frac{\chi(\cos k_x +\cos
k_y)^{2}}{E_k},
\\
\frac{\partial E_{mf}}{\partial \Delta}&=& 2 \Delta  = \sum_k \frac{\Delta(\cos k_x -\cos k_y
-a_0^1)(\cos k_x -\cos k_y)}{E_k}, \\
 a^1_{0} &=& 0.
\end{eqnarray}
where the dispersion $E_k$ is given by
\begin{eqnarray}
E_k = \sqrt{(\chi_k - \anot^3)^2 + (\Delta_k - a^1_{0})^2 }, \nonumber \\
\chi_k = -\frac{3J}{4}(\cos k_x + \cos k_y)\chi, \nonumber \\
\Delta_k = -\frac{3J}{4}(\cos k_x - \cos k_y)\Delta.
\end{eqnarray}
In the empty limit, the holons cannot move, so there are no mean field equations and
dispersions governing those. The above mean field equations can be solved numerically to yield
the unphysical result $\chi=\Delta=\frac{\sqrt{2}}{4}$, identical to the result for half
filling. But this is clearly nonsense: the empty limit is neither a spin liquid nor a
superconductor. Also, since $\Delta$ and $\chi$ are non-zero, the total energy will be
nonzero, in flagrant contrast with the correct result being zero, as pointed out earlier.

Taking $a^1_{0}$ into account, however, the above mean field equations are extended with the
saddle point equation for $a^1_0$,
\begin{equation}
1 = \sum_k \frac{(\Delta(\cos k_x -\cos k_y) -a^{1}_0)}{E_k},
\end{equation}
where the number 1 is the boson density. Solving the new system of equations numerically, we
obtain the correct result $\chi=\Delta=0$ and $a_0^1=\frac{1}{2}$. Therefore, the Lagrange
multiplier is of central importance, and the mean fields vanish, as they should. Substituting
this solution  in the Hamiltonian Eq.(\ref{mftheory}), we recover the correct energy for the
empty limit. In other words, things go dramatically wrong if $a^1_0$ is ignored. In Section
\ref{sect:psds} we will show that our mean field theory is performing well as an energy
density functional at intermediate dopings. To calculate dynamic properties, one has to be
careful in choosing the correct mean field approach.

\subsection{Dynamical properties of the empty limit}\label{subsect:dynempty}

It is interesting to consider the dynamic properties following from the slave boson theory in
the empty limit. It turns out that although this theory is a good energy density functional,
it is less trustworthy with regard to dynamical properties revealed through the propagators.
This is of course due to the mean field treatment ignoring  fluctuations of the gauge fields.
Indeed, the Lagrange multipliers $a^l_{0i}$ should be given dynamics, causing fluctuations
confinement of the spinons and holons to electrons. Since we ignored the fluctuations, we can
not expect the mean field theory to describe the physical electron of the empty limit.

The starting point for the study of electron dynamics is the single electron propagator
\begin{eqnarray}
G(x,y;t-t') &=& \langle T(c_{x\uparrow}(t)c^{\dag}_{y\uparrow}(t'))  \rangle  \\
&=& \langle c_{x\uparrow} e^{-i\hat{H}(t-t')/\hbar} c^{\dag}_{y\uparrow} \rangle\Theta(t-t')\nonumber \\
& &  - \langle \langle c^{\dag}_{y\uparrow} e^{-i\hat{H}(t'-t)/\hbar}
c_{x\uparrow}\rangle\Theta(t'-t). \nonumber
\end{eqnarray}
Here the $c_{i\sigma}$ again describe projected electron operators. Since in the empty limit
there are by definition no electrons in the vacuum,
\begin{eqnarray}
G(x,y;t-t') &=& \sum_{kk'} e^{ik'x-iky} \langle
c_{k'\uparrow}e^{-i\hat{H}(t-t')/\hbar}c^{\dag}_{k\uparrow} \rangle \nonumber \\
& & \times \Theta(t-t')e^{i\hat{E}_0(t-t')/\hbar}.
\end{eqnarray}

Let us first show that for the exact expression (\ref{emptyexact}) describing the  empty limit
of the $t-J$ model, one obtains a free particle dispersion.

 We need to know
the time evolution operator $e^{-i\hat{H}(t-t')/\hbar}$. The Hamiltonian operator has no
Heisenberg part for projected electrons, and it also vanishes on the state
$c^{\dag}_{x\uparrow} | 0\rangle $, where $| 0\rangle $ is the wave function
Eq.(\ref{emptyexact}). So we only need to calculate the effect of
$H_t=-t\sum_{ij}\psi_i^{\dag}h_i h_j^{\dag}\psi_j$ on
\begin{equation}
c^{\dag}_{x\uparrow} | \mbox{empty}\rangle = \frac{1}{V}\sum_m
e^{imx}(b_{1m}f^{\dag}_{\uparrow m}+b_{2m}f_{\downarrow m}) | 0\rangle.
\end{equation}
In Fourier space, the result is
\begin{equation}
\langle c_{k'\uparrow} H_t c^{\dag}_{k\uparrow} \rangle = -2t\delta_{kk'} (\cos k_x + \cos
k_y) \equiv \varepsilon_k \delta_{kk'}.
\end{equation}
Including the chemical potential and using a contour integral expression for $\Theta(t-t')$,
we conclude that the propagator is
\begin{equation}
G_{exact}(k,\omega) = \frac{1}{\hbar\omega - (\varepsilon_k -\mu) +i\eta},
\end{equation}
the correct result for the propagator of a free particle. This means that the wave function
Eq.(\ref{emptyexact}) encodes the right physics, as expected.

What performance can be expected from the mean field theory when asked dynamical questions?
Let us first consider the case for the mean field Hamiltonian Eq. (\ref{mftheory}). Without
$a^0_1$, one obtains the expected errors, namely a $d$-wave dispersion. Repairing this with a
nonzero $a^1_0$ leads, however, to just partially good news. We showed that the $d$-wave
dispersion vanishes, which is physical. The bad news, however, is that both $\chi$ and
$\Delta$ are zero. This absence of kinematics would lead to a dispersionless spinon spectrum.
This is not what one would expect, since shooting electrons into the void should behave as a
free electron cosine dispersion.

This motivates the second approach: let us apply only mean-field theory at the wavefunction
level, \emph{without} introducing the Hubbard-Stratonovich fields Eq.(\ref{chi}) and
Eq.(\ref{Delta}). This means that the Hamiltonian then reads

\begin{eqnarray}\label{hexact}
H_{exact} &=& -\mu \sum_{i} h^{\dag}_{i}h_{i} - \sum_{i} a_{0i}^{l}
\left(\frac{1}{2}\psi^{\dag}_{\alpha
i}\tau^{l}\psi_{\alpha i} + h^{\dag}_{i}\tau^{l}h_{i} \right) \nonumber \\
& & +  \sum_{< ij >} \frac{3J}{8} \left( \mathbf{S}_i \cdot \mathbf{S}_j \right) - \sum_{<ij>}
t\left( \psi^{\dag}_{i} h_i h_j \psi_{j} + h.c. \right). \label{exacttheory}
\end{eqnarray}

In the empty limit, we just have one contribution to the propagator
\begin{eqnarray}
G_{mf}(x,y;t-t') &=& \sum_{kk'} e^{ik'x-iky} \langle
c_{k'\uparrow}e^{-i\hat{H}(t-t')/\hbar}c^{\dag}_{k\uparrow} \rangle \nonumber \\
& & \times \Theta(t-t').
\end{eqnarray}
The expectation values are taken with respect to the empty-limit mean field state
Eq.(\ref{emptyMFstate}), denoted by $\left| 0 \right> _{MF}$. It is convenient to first
calculate how the electron operator acts on the empty limit mean field state,
\begin{eqnarray}
c^{\dag}_{x\su}\left| 0\right>_{MF} &=& \frac{1}{\sqrt{2}}( b_{1x}f^{\dag}_{x\su} +
b_{1x}f_{x\sd} ) \nonumber \\
& & \times \prod_{l} \half (b^{\dag}_{1l} + b^{\dag}_{2l})(1 +
f^{\dag}_{l\sd}f^{\dag}_{l\su}) \vacket \nonumber \\
&=& \frac{1}{\sqrt{2}} f^{\dag}_{x\su} P_{x} \vacket,
\end{eqnarray}
with the definition $P_x \equiv \prod_{l\not= x} \half (b^{\dag}_{1l} + b^{\dag}_{2l})(1 +
f^{\dag}_{l\sd}f^{\dag}_{l\su})$.

First we demonstrate that the state $c^{\dag}_{x\su}\left| 0\right>_{MF}$ is a physical state,
i.e., it satisfies the constraint Eq.(\ref{mfconstr}). It does, since $b_{\alpha x}P_x\vacket
= 0 $ makes the expectation value $\langle h^{\dag}_{x}\tau^{l}h_x \rangle $ zero.
Furthermore, the fermionic expectation values vanish as well. As an example,
\begin{eqnarray}
& & \left< 0\right| c_{y\su} \psi^{\dag}_{y}\tau^{1} \psi_{x} c^{\dag}_{x\sup} \left| 0
\right> _{MF} = \half \left< 0\right| P^{\dag}_{y} f_{y\su}f_{x\sd}P_{x}  \left| 0
\right>_{MF}
\nonumber \\
 &=&
-\half\left< 0\right| P^{\dag}_{y} f_{y\sd}f_{x\sd}P_{x,y}  \left| 0 \right>_{MF}=0,
\end{eqnarray}
both for $x=y$ and $x\not=y$. The other $l$'s are checked similarly. This implies that the
constraint terms in the Hamiltonian Eq.(\ref{hexact}) vanish.

Then we demonstrate that the Heisenberg term also vanishes on $c^{\dag}_{x\su}\left|
0\right>_{MF}$. This can be seen easily since $\mathbf{S}_{i}\cdot \mathbf{S}_{j}
f^{\dag}_{j\su}P_{j}\vacket = \mathbf{S}_{j} f^{\dag}_{j\su} \cdot \mathbf{S}_{i} P_{j}\vacket
= 0$, by the same reason why the Heisenberg term of the mean field energy on the empty state
vanishes.

Only the hopping terms are non-trivial. This is intuitively clear, since shooting an electron
in the empty sample, removes one holon, allowing the rest to move. This is corroborated by
calculating the matrix elements of the \emph{exact} hopping Hamiltonian $H_t = -t\sum_{\langle
ij \rangle} \psi^{\dag}_{i} h_i h_j \psi_{j}$:
\begin{eqnarray}
& & \left< 0 \right| c_{y\su} H_t c^{\dag}_{x\su} \left| 0 \right>_{MF} \nonumber \\
&=& \sum_{ij} \vacbra P^{\dag}_{y} (b_{1y} + b_{2y}f_{y\su}f_{y\sd})(b^{\dag}_{1x} +
b^{\dag}_{2x}f^{\dag}_{x\sd}f^{\dag}_{x\su}) P{x}\vacket \nonumber \\
&=& \frac{1}{4} \sum_{\langle yx \rangle} \vacbra (1 +
f_{x\su}f_{x\sd})(f^{\dag}_{y\sd}f^{\dag}_{y\su}) \vacket \nonumber \\
&=& \sum_{\langle ij \rangle} \delta_{ix}\delta_{jy},
\end{eqnarray}
meaning that the electron can only hop to a nearest neighbour site if that site is empty. The
Fourier transform then reads
\begin{equation}
\left< 0\right| c_{k'\su} H_t c^{\dag}_{k\su} \left| 0 \right>_{MF} = -\half t \delta_{k'k}
(\cos k_x + \cos k_y) = \varepsilon_k \delta_{k'k} ,
\end{equation}
the free particle dispersion. This leads to the same free particle propagator as in the exact
case,
\begin{equation}
G_{mf}(k,\omega) = \frac{1}{\hbar\omega - (\varepsilon_k -\mu) +i\eta}.
\end{equation}

In other words, when one treats the slave boson Hamiltonian exactly, the mean field wave
function for the empty limit gives the correct single-electron propagator.

\section{The phase separated $d+s$-wave superconductor } \label{sect:psds}

In the previous section, we have discussed how one should describe doping within the context
of $SU(2)$ gauge theory. By studying this carefully, we convinced ourselves of two important
lessons. The first one is that a hole in the $t-J$ model is described in the $SU(2)$ theory by
a superposition of a vacancy and a spin pair singlet state. This implied that doping induces
$s$-wave pairing. The second message is that the holons describing doping should be hard core
bosons, instead of gaseous, non-interacting particles employed by Wen and Lee. This hard-core
is necessary condition in order to account for the fact that there can be at most one charge
per site, as has been clear from very the beginning.

Our results are summarised in the mean-field phase diagram, reflecting the phase separated
$d+s$-wave superconductor. Another ramification we make quantitative. The $s$-wave admixture
in the superconducting gap is shown to shift the gap nodes along the Fermi surface. We predict
how this node shift behaves as a function of doping, an effect which might be just within the
resolution of present day angle resolved photo-emission experiments.

At the end of this section, we will discuss the actual meaning of the mean field states of
$SU(2)$ mean field theory, with respect to the question which mean field states are
superconductors, and which are not. The distinction will be made by the absence or presence of
the Meissner effect.

\subsection{$SU(2)$ energy density functional}\label{subsect:edf}

In the previous sections, we already set the stage for the mean field theory description of
the doped Mott insulator. Now we are in the position to derive the energy density functional.
Inspired by the empty limit, our starting point is the mean field wave function,
\begin{equation}
| \Psi_0 \rangle_{MF} = \prod_i \left( \alpha_i + \beta_{i}e^{i\varphi_i}(u_i  b^{\dagger}_{i1
} + v_i b^{\dagger}_{i2 } ) \right) \vacket
 \left|\mbox{F}\right> \label{MFstate}.
\end{equation}
The ket $\left| F \right>$ describes the many body spinon state, and the boson density
$|\beta|^2$ is the density of physical holes.

The important point of $SU(2)$ gauge theory is that the particle-hole symmetry is broken upon
doping. Indeed, a hole is described by a superposition of vacancies and spin pair singlets
$\fdag_{i\sd}\fdag_{i\su}$, accompanied by their own boson. In the particle-hole symmetric
state, the $b_1$ and $b_2$ boson are equal, meaning that this should correspond with
$\theta=\half\pi$. This is the motivation for the parametrisation $u_i = \cos(\half\theta_i)$
and $v_i = \sin(\half\theta_i)$.

The phase $\varphi$ is the same as the phase $\varphi_i$ of $g_i$, which is gauge. From now
on, we gauge $\varphi_i=0$ everywhere. The transformation (\ref{gispecial}) mapping the SFP
into the dSC corresponds with $\varphi_i = \half \pi + I \pi$, as the reader can verify.

For theoreticians, it is natural to first study spatially homogeneous mean field states.
However, in the course of time it has become clear that strongly interacting electron systems
tend to form inhomogeneous states, like stripes. Being aware of this complication, let us
nevertheless study homogeneous states. Still, this exercise turns out to be instructive, in
this regard. The reason is that we treat the Hamiltonian Eq. (\ref{mftheory}) in the grand
canonical ensemble, instead of the canonical ensemble. The original formulation of  the
$SU(2)$ mean field theory  \cite{wenlee, lnw, lnwng} rested on the canonical ensemble as well.
To account for condensation of the holons at finite doping,
 the temperature was taken to be finite, to find out that the particle number constraint leads to
Bose-Einstein condensation of the holons, by treating $\mu$ simply as a Lagrange multiplier.
We prefer a different approach, since considering first finite temperature is a detour given
in by the unphysical assumption htat the holons form a non-interacting gas. On the other hand,
hard-core bosons are not only more physical, but also easy to treat at zero temperature. In
Section \ref{sect:empty} we made the point that the bosons are interacting, making them
superfluid at zero temperature.  The last  motive is the possibility of phase separation,
i.e., the possibility of coexistence of phases with different densities at the same chemical
potential in the same volume, giving rise to the need of performing the Maxwell construction.
To anticipate this, it is necessary to take the chemical potential for the holons as control
parameter, instead as the Lagrange multiplier enforcing a fixed density.

Let us substitute the mean field wave function Eq. (\ref{MFstate}) in the slave Hamiltonian
Eq.(\ref{mftheory}), we obtain an expression for the mean field energy per site
$\frac{1}{N}\left< H_{mf} \right> = e_{MF}$. ($N$ is the number of lattice sites.)
\begin{eqnarray}
e_{MF} & = & -\frac{1}{N} \sum_k E_k +  \frac{3}{4 N}J( |\chi|^{2} + |\Delta|^{2})
\\
& & - \; 2 t \chi |\alpha|^2 |\beta|^2 \nonumber - (\mu + a_0^1 \sin \theta + a^3_0 \cos\theta
) |\beta|^2.
 \label{mfielden}
\end{eqnarray}
The homogeneity of the Ansatz is expressed in the fact that the lattice site subscript $i$ is
dropped. It is important to observe that the isospin latitude angle appears in the mean-field
energy, expressing the fact that for non-zero doping, $SU(2)$ gauge symmetry is broken. Hence
 $\theta$ is not gauge, but has acquired physical meaning. The kinetic part of the
energy is gauge invariant, leading to the fact that only the hopping amplitude shows up in the
holon hoppings. From this density functional, we derive the saddle point equations for the
dynamical variables $\chi, \Delta, a^1_0, a^3_0$ and the hole density $|\beta|^2 = \rho(\chi)
= 1 -|\alpha|^2$. To simplify matters a  bit, we take the isospin angle $\theta$ as an
external parameter, controlling the density of $b_2$ relative to $b_1$.

The saddle point equations in the grand canonical ensemble are easily derived, with the
homogeneous forms of Eq.(\ref{constr1}) and Eq.(\ref{constr3}),
\begin{eqnarray} \label{mfeq} 2
\chi &=& \frac{1}{\chi} \sum_k \frac{\chi_k (\chi_k -a^3_0)}{\ E_k}+2\left(\frac{4 t}{3 J}
\frac{\partial}{\partial \chi}
 \rho(\chi) (1 -\rho(\chi)\ ) \right) \nonumber
\\
2 \Delta  &=& \frac{1}{\Delta}\sum_k \frac{\Delta_k(\Delta_k -a^{1}_0)}{ \ E_k} \nonumber
\\
\rho(\chi)\sin(\theta) &=& \sum_k \frac{(\Delta_k -a^{1}_0)}{E_k} \nonumber
\\
\rho(\chi)\cos(\theta) &=& \sum_k \frac{(\chi_k -a^{3}_0)}{E_k} \nonumber
\\
0 &=& \rho(\chi)\left( \rho(\chi) - \frac{1}{2}\left( 1 + \frac{\mu + a^{1}_{0}\sin\theta +
a^{3}_{0}\cos\theta}{2t\chi} \right)  \right). \nonumber
\end{eqnarray}

As already discussed after Eq.(\ref{constr1}), these equations give rise to an important law
which provides a linear relationship between the $s$-wave spinon pairing $\Delta_s \equiv
\left< \fdag_{i\sd}\fdag_{i\su} \right>$ and the doping $x = \rho(\chi)$,
\begin{equation}\label{swavedoping}
\Delta_s = \half x \sin(\theta),
\end{equation}
as a direct ramification of the full constraint structure.

In order to find the solutions to the saddle point equations Eq.(\ref{mfeq}),  the energy
Eq.(\ref{mfielden})  is minimised numerically using the simulated annealing method
\cite{simann,numericalrecipes}.

We point out that the fifth equation admits both zero and non-zero solutions for the density
$\rho$. As a function of $\mu$, the mean field energy will tell which one is more favourable.
In  Section \ref{subsect:mfdiag}, we will show that the system chooses between these two by a
first order phase transition, and not a second order one! This means that the mean field
theory \ref{mfielden} implies a phase separation regime, for the usual Maxwell construction
reasons, when transforming to the canonical ensemble.

\subsection{The $SU(2)$ mean field phase diagram} \label{subsect:mfdiag}

We have now arrived at the point where we can collect the results. Our  first step was to
prove that one needs all $SU(2)$ constraints to project onto the $t-J$ model Hilbert space,
while this constraint structure is also required for the mean field description of the $SU(2)$
gauge theory. This will bring us to the first result: that the superconducting order parameter
\textit{needs} to have an $s$-wave component! Then we spent effort  in proving that the holons
need to be hard-core in order to respect  the full set of $SU(2)$ constraints.  We now show
that the superfluid hard-core holon condensate displays phase separation behaviour, as
expected for hard-core interacting systems. We thereby achieve an intrinsic connection between
slave theories on the one hand, and the observation of inhomogeneous states on the other hand,
a connection that is traditionally considered as absent.

Let us first discuss the numerical verification of a claim inferred in the literature
\cite{wenlee}, namely that at finite hole density the superconducting state ($\theta =\pi/2$)
is preferred over the flux phase ($\theta = 0, \pi$) (cf. the inset in Fig. 1). This is a
natural ramification of the breaking of $SU(2)$-symmetry for non-zero doping.
The $\theta= \frac{\pi}{2}$ states, characterised by $a_0^3=0$, are energetically more
favourable, being consistent with the instability of the flux state towards $d$-wave
superconductivity, as already understood in the early nineties \cite{kotliar}.

\begin{figure}
\label{mushift}
\centering \rotatebox{0}{}
\resizebox{7.3cm}{!}{%
\includegraphics*{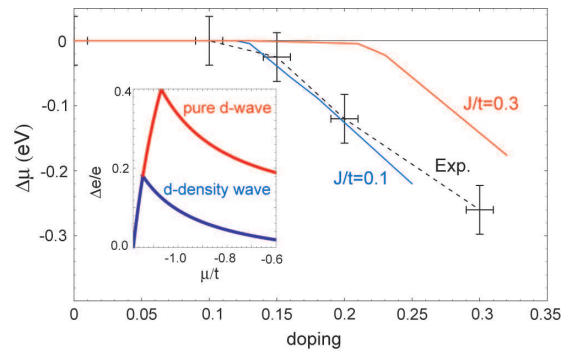}}
\caption{ \small{Chemical potential shift as a function of doping, showing the phase
separation behaviour of the reformulated mean field theory. Indeed, the chemical potential
starts to shift for appreciable dopings only. The blue line (colour online) are the numerical
results for $J/t =0.1$, and agrees very well with the experimental results from Fujimori
\cite{fujimori} (dotted line). The red line are the results for $J/t=0.3$. The critical doping
changes, but not the compressibility. The inset shows that ignoring $a^1_0$, i.e., ignoring
the $s$-wave component, gives a false vacuum. Indeed, there is a positive relative energy
difference between the pure dSC and  our mean field theory, growing with doping (lower line).
 }}
\end{figure}

The reader might already have noticed a kink in $e_{MF}$  as a function of doping in the inset
figure \ref{mushift}. In other words: there is a first-order phase transition. Indeed, our
main result is that generically this mean-field theory predicts phase separation at small
chemical potential. The system stays initially at half-filling and pending the ratio of $J/t$
at some finite $\mu$ a level crossing takes place to a state with a finite doping level, cf.
Fig.1.

We stress here that this phase separation behaviour is eventually coming from the hard-core
nature of the holons:  also the "wrong" mean field states (SFP and pure dSC with pockets)
exhibit first order behaviour.  We conclude that, due  to the Hilbert space restrictions on
the $SU(2)$ description of the holons, the theory insists on inhomogeneous states for low
doping. As a function of increasing $J/t$ the width of this phase separation regime is
increasing (see Fig. 2) and we find that for $J/t \simeq 4$ the phase separation is complete.
This is consistent with exact diagonalisation studies on the t-J model indicating a complete
phase separation for $J/t \geq 3.5$ \cite{kivelsonPS},\cite{hellberg}. This is quite
remarkable and it reveals that the gauge mean field theory has to be a remarkably accurate
quantitative theory of the density functional kind: it is a good description of the empty
limit and the Mott insulator, and gives a fair prediction of phase separation tendencies. We
stress that as a rule less severe demands on physical reality are put on density functional
theory , instead of full dynamical theories.

\begin{figure}
\centering \rotatebox{0}{}
\resizebox{5.3cm}{!}{%
\includegraphics*{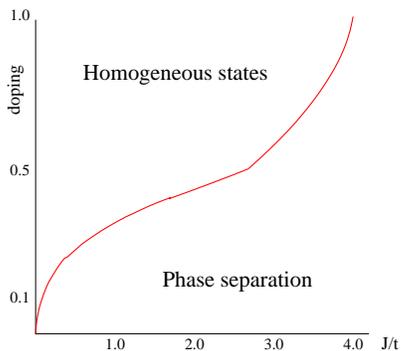}}
\caption{ \small{Phase diagram as a function of doping $x$ and the ratio $J/t$, according to
the Maxwell construction. For dopings below a critical doping line $x_c(J/t)$, homogeneous
states are metastable against phase separation, stripes etc. For $J \simeq 4t$, total phase
separation takes place and the system becomes a mixture of Mott insulating and empty regions.
  }} \label{fuji}
\end{figure}

The  significance of this finding is that for this most sophisticated version of spin-charge
separation theories, phase separation is natural feature, as it is in the empirical reality.
 It
is well understood that these macroscopic phase separated states are an artefact of the
oversimplified $t-J$ model. By taking the long-range Coulomb interaction into account this
will turn immediately into the microscopic inhomogeneity
\cite{zaanen},\cite{white},\cite{kivelsonstripe}, of the kind that are seen in STM-experiments
\cite{seamus}. To see how well this slave theory handles the 'big numbers' in this regard, we
show in Fig. 1 the electronic incompressibility $1 /\kappa =
\frac{\partial^{2}E_{\mbox{\scriptsize{MF}}}}{\partial x^{2}} = \frac{\partial\mu}{\partial x
}$ according to the $SU(2)$ theory, to find that it compares remarkably well with the
experimental results due to Fujimori and coworkers\cite{fujimori}. Firstly, it is seen that
independent of the ratio $J/t$, the compressibility is right on spot of the experiments: the
slope of $\mu$ vs. holon density is the same as the measured slope. The doping at which phase
separation occurs, namely 13 \%, is correct only for the value $J/t=0.1$, which is too small.
Indeed, from ARPES measurements a ratio of $J/t=0.3$ is more realistic, but for those samples
phase separation takes place at dopings of about 17 \%, and not 21 \%, as found in the $SU(2)$
mean field theory for $J/t=0.3$.

Let us now focus on the nature of the superconducting order parameter found  elevated doping
levels. As expected, the $s$-wave component becomes increasingly important, cf. Eq.
(\ref{swavedoping}). To further emphasize this , we compare in the inset of Fig. 1 the energy
of a state where we have fixed the Lagrange multiplier $a^{1}_{0}=0$ such that the s-wave
component vanishes, with the best $d+s$ mean-field state, finding that the former is indeed a
false vacuum. To mimick the average behavior of the superconducting order parameter also in
the micro-phase separated states at low dopings, we calculate matters now in the false
(uniform) vacuum of the canonical ensemble, fixing the average density, simplifying the mean
field equations Eq.(\ref{mfeq}). Indeed, $\rho(\chi)$ becomes now a fixed $\rho$. Furthermore,
since the state $\theta=\half\pi$ is favoured, we take the mean field Ansatz $U_{ij}$ to be
the SC one, as stated before, and consequently we have $a^3_0=0$. This enables us to map out
the phase diagram as a function of doping and the ratio $J/t$, leading to three phases. The
first, for low doping, is the phase-separated, underdoped $d+s$-wave superconductor. It is a
mixture of charged, superconducting islands in an insulating sea without charges, where  the
full $SU(2)$ symmetry is restored. This reminds the reader of the STM-pictures from S.C.
Davis' group \cite{seamus}. For intermediate dopings, the homogeneous, overdoped $d+s$-wave
superconductor is found, whereas for high dopings, the $d$-wave gap vanishes, leaving behind a
pure $s$-wave superconductor. Hence, although $d$-wave superconductivity leads to an $s$-wave
admixture, the reverse is not true.

\begin{figure}
\centering \rotatebox{0}{
\resizebox{7.3cm}{!}{%
\includegraphics*{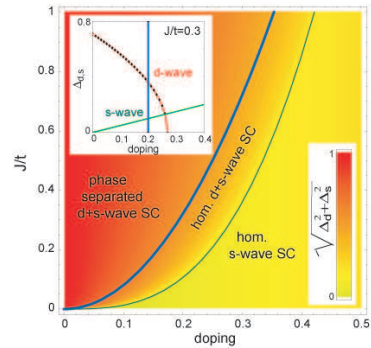}}}
\caption{ \small{ Zero temperature phase diagram of the $t-J$ model according to $SU(2)$-mean
field theory. It incorporates three phases, viz. the phase separated $d+s$-wave superconductor
for low dopings, the homogeneous $d+s$-wave superconductor for intermediate dopings, and a
homogeneous $s$-wave superconductor at high dopings. The bold line shows indicates the border
of the phase separation region. The phase separation tendency grows for increasing $J/t$, to
become complete at $J/t=4$. The colors indicate the total superconducting gap. For zero
doping, there is only a $d$-wave component, whereas the $s$-wave admixture grows linearly with
doping, so that the total gap is non-zero even beyond the critical $x-J/t$ line (rightmost
line), where $\Delta_d$ vanishes. The inset shows the $d$-wave and the growth of the $s$-wave
component separately for $J/t=0.3$. The blue line indicates the doping level where phase
separation terminates, computed by imposing uniformity (canonical ensemble) for
$x<0.2$.}}\label{phasediag}
\end{figure}

We find that the regime where phase separation is important, the $s$-wave component is  not
negligible. This is consistent with Raman measurements \cite{tajima}, where the
superconducting gap was found to have both $d$- and $s$-wave components. Although screening
effects in Raman scattering make it difficult to compare our results directly to theirs, their
results indicate that the ratio $r= \Delta_s/\Delta_d  $ grows with doping, as it does in our
approach. Looking to Figure 3.,
 we find in the phase separated overdoped regime  $s$-wave admixtures
of about $r=10-20 \% $,  consistent with $c$-axis tunneling experiments \cite{klemm2005}.

We predict that at a doping level that appears to be higher than can be achieved in cuprate
crystals a phase transition occurs to a pure $s$-wave superconductor. As we already alluded
to, the gauge fluctuations should become more severe as well, for increasing doping and at
some doping level a transition should occur to a confining "electron like" system.

As we learned from the empty limit, it still make more sense than the result obtained by
disregarding the first constraint equation Eq.(\ref{constr1}), since that would lead to the
unphysical result $\chi=\Delta=\frac{1}{\sqrt{2}}$, giving the wrong vacuum energy, as we
discussed in Section \ref{sect:empty}.  In other words, our mean field theory is a remarkably
good density functional theory, but inevitably the theory fails completely in dynamic regards.
In the confined phase, at sufficiently high dopings, we need an approach which is
qualitatively different from slave theories, let alone that one can get away with the mean
field version. On the other hand, in the superconducting doping regime, we find some promising
experimental support for our results with regard to the $d+s$ structure of the order
parameter, meaning that confinement physics might not be overwhelmingly important in the
low-doping part of the phase diagram.

Having said this, the experimental support for an $s$-wave admixture makes it possible to come
up with a falsifiable prediction for photo-emission experiments. The $s$-wave component
induces a shift in the nodes, as can be inferred from the spinon dispersion $E_k$,
Eq.(\ref{Ek}): the hopping vanishes along the line $k_y = \pi -k_x$, so that the locus of the
node can be readily calculated to be $k_y  - \half \pi=  \arccos(a^1_0/ (3J/2)\Delta)$, which
shows a doping-dependent behaviour as well, cf. the blue line in Figure 3. The node shifts
might be able to explain the U-shaped gap as measured in Bi2212 \cite{borisenko}, since the
twinning of samples mixes regions of node shifts $+\delta$ with $-\delta$, smearing out the
V-shape of the gap to a U-shape.

\begin{figure}
\centering \rotatebox{0}{
\resizebox{7.3cm}{!}{%
\includegraphics*{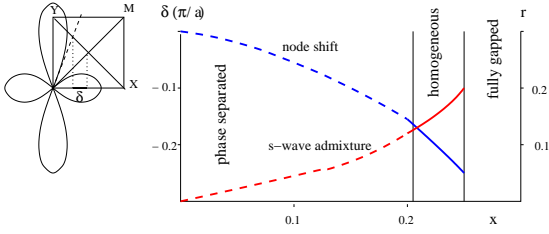}}}
\caption{\small{ In this figure, we show how the $s$-wave admixture  $r=\Delta_s/\Delta_d$
grows as function of doping (red line) for $J=0.3t$. The insets show the symmetry of the gap
function in reciprocal space, indicating the shift of the gap node along the Fermi surface,
coinciding with the Brillouin zone boundary Y-X. This shift is denoted by $\delta$, and grows
with doping. The blue line plots $\delta$ as function of doping in units of $\pi/a$, where $a$
is the lattice spacing. Up to a doping of 20\% the results of the false vacuum homogeneous
solution in the canonical ensemble are used, exploiting the Maxwell construction. To remind
the reader of the fact that the homogeneous states are false vacuum states, the lines are
dashed. Finally, the lines end where the $d$-wave gap vanishes. Here $r$ is infinite and the
Fermi surface is fully gapped. }}\label{nodeshift}
\end{figure}

\subsection{Single-electron dynamics}\label{subsect:electdyn}

It is also possible to interrogate the single-electron dynamics of the superconducting state.
Wen, Lee, Nagaosa and Ng have addressed this issue for non-interacting holons \cite{lnwng}.
Since we showed that one cannot ignore the hard-core nature of the holons for appreciable
dopings, it is interesting to investigate what the electron correlation function looks like.
We will demonstrate that it consists of two parts, ubiquitous in hard-core interacting Bose
systems \cite{fetterwalecka}: one is a coherent condensate piece, and the second is an
incoherent piece displaying sound modes for long wavelengths. This novel result is entirely
due to the hard-core nature of the bosons, and might lead to falsifiable experimental
predictions.

The calculation can be a bit simplified by the fact that in the superconducting state $\theta
= \half \pi$, hence we write $b_1 = b_2 = b$. Then the electron propagator
\begin{equation}
G(k,t-t') = \langle\langle c_k(t)c^{\dag}_k(t') \rangle\rangle
\end{equation}
is the convolution of bosonic operators $b_{p}$ and $f_{k+p}$, which can be seen by
Fourier transforming the expression for the electron $c_{i\su} = b^{\dag}_{i1}f_{i\su} +
b^{\dag}_{i2}f^{\dag}_{i\sd}$.
The time resolved propagator  $ G(k,\omega)$ can be calculated by using the spinon coherence
factors $u_{k+p}$ and $v_{k+p}$ from the Bogoliubov diagonalisation of the spinons
\begin{eqnarray}
f_{q\su} &=& u_{q}\gamma_{q0} + v_{q}\gamma^{\dag}_{q1} \\
f^{\dag}_{q\sd} &=& -v_{q}\gamma_{q0} + u_{q}\gamma^{\dag}_{q1}
\end{eqnarray}
Exploiting the Fourier transformation of the textbook result \cite{fetterwalecka} for the
boson propagator, the frequency resolved propagator then reads for finite repulsion $U$
\begin{widetext}
\begin{eqnarray}
G(k,\omega) = \sum_{p} \int \frac{d\Omega}{2\pi i}  & & \left\{ \frac{-\left| u_{k+p} -
v_{k+p}\right|}{(\hbar\omega-\hbar\Omega + \varepsilon^{f}_{k+p}) + i\eta} \left[
x_{0}\delta(p)\delta(\Omega) - \frac{\hbar(\Omega+\Omega_p)}{\hbar^{2}(\Omega_p -\Omega)^{2}
-(2Ux_{0})^{2}} \right]
\right. \nonumber \\
& & + \left. \frac{\left| u_{k+p} + v_{k+p}\right|}{(\hbar\omega-\hbar\Omega +
\varepsilon^{f}_{k+p}) - i\eta}\left[ x_{0}\delta(p)\delta(\Omega) +
\frac{\hbar(\Omega+\Omega_p)}{\hbar^{2}(\Omega_p + \Omega)^{2} -(2Ux_{0})^{2}} \right]
\right\}.       \label{convolprop}
\end{eqnarray}
\end{widetext}
Here, the fermion and boson dispersions are denoted by $\varepsilon^{f}_{k}$ and
$\hbar\Omega_{p}$, respectively. Although the expression for the propagator
Eq.(\ref{convolprop}) seems lengthy at first, it conveys a couple of important messages. The
most obvious one is the fact that the propagator is a convolution of bosons and fermions. The
second one is the separation between the condensate piece, proportional to $x_0$, and an
incoherent piece. The condensate piece leads to poles at the fermion dispersions, $\hbar\omega
= \varepsilon^{f}_k$. The incoherent piece is even more interesting. Since for the bosons
$\hbar\Omega_p = 2t\chi(\cos k_x + \cos k_y)$, we have for the long wavelength limit and for
$\mu\simeq 2t$ (the requirement for a condensate to exist) that the holon poles are located at
\begin{equation} \label{finiteUomega}
\hbar\Omega \simeq 4t\chi\sqrt{2Ux_0} |k|  \mbox{\ \ \ \  finite \ } U.
\end{equation}
The holon velocity of sound can be written in a form for hard-core bosons as well, since for
finite $U$, $x = \frac{1}{2U}(2t\chi+\mu)$, whereas for hard core bosons $2t\chi + \mu =
2x_0$. Hence the result Eq. (\ref{finiteUomega}) can be generalised to infinite repulsion $U$,
\begin{equation}
\hbar\Omega = 2 (2t)^2 \sqrt{x_0} |k| \mbox{\ \ \ \ infinite \ } U.
\end{equation}
This means that for the incoherent piece of the propagator Eq. (\ref{convolprop}), the poles
are located at
\begin{equation}
\hbar\omega = \varepsilon^{f}_{k+p} - v|k|, \mbox{\  \ \ } k \rightarrow 0.
\end{equation}
with the velocity of sound $v= 2 (2t)^2 \sqrt{x_0}$.

In summary, we have argued that in order to achieve consistency in the $SU(2)$ slave boson
theory, one has to implement hard-core bosons instead of gaseous Bogoliubov bosons to describe
the charge sector. As a result, one obtains phase separation at lower dopings consistent with
the experimental observations. Also the compressibility matches very well. Furthermore, by
inspecting  the empty limit, we showed that  an $s$-wave component in the superconducting
order parameter is implied  when $d$-wave superconductivity occurs, at least for dopings where
homogeneous states exist, a ramification of the constraint structure. This finds its origin
eventually in the particle-hole symmetry central to the gauge structure of the $SU(2)$ theory:
to describe physical spin singlets,``no fermions" are indistinguishable from an ``s-wave
spinon pair". This is reflected in the constraint equations, required to reduce the $SU(2)$
Hilbert space to the Hilbert space of the $t-J$model. The constraint equations tell us that as
soon as $d$-wave superconductivity emerges, one necessarily has an s-wave component.  This
s-wave admixture is in accord with Raman \cite{tajima} and $c$-axis tunneling experiments
\cite{klemm2005,klemm2006}. We also predict a node shift in the gap function, that might be
measurable by photoemission.

\section{Isospin spiral states in cuprates} \label{sect:isospin}

Having established the phase separation tendencies, new perspectives are opened as to what the
role of stripes in the superconducting cuprates is. The understanding is that phase separation
is a necessary condition for stripes to exist. On the other hand, phase separation has never
been seen in the high-$T_c$'s. In fact, there are other properties at work. The Mott insulator
is an antiferromagnet, which makes it advantageous for holes to order in stripes instead of
phase separated islands. In fact, stripes should be regarded as antiphase boundaries in the
antiferromagnet \cite{zaanengeomorder}.

In the past decades, many approaches to understand the Hubbard model in some slave boson
representation are made.
 One is the large-$S$ expansion \cite{haldane,assa},
taking the limit of the spin value $S\rightarrow \infty$. The other is introducing more than
two flavours of spin, such that an $SU(N)$ model is obtained . The large-$N$ limit leads to
dimerised states \cite{readsachdev,affleck3}, whereas the vacuum of the large-$S$ limit is the
antiferromagnet \cite{haldane, assa}. The question arises if the group $SU(2)$ is able to
describe the ``anti-phase boundariness'' of the antiferromagnet, the more so since the
antiferromagnet and the dimerised state are incompressible, whereas the flux phases and the
dSC spin liquids of the $SU(2)$ theory are compressible.

\begin{figure}
\centering \rotatebox{0}{
\resizebox{7.3cm}{!}{%
\includegraphics*[width=7.3cm, height=4.3cm]{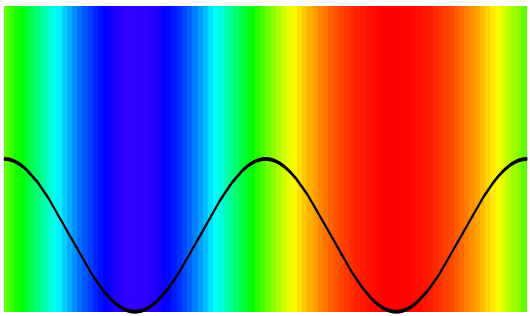}}}
\caption{\small{ A pictorial representation of the unit cell of the isospin spiral. The green
areas correspond with superconducting stripe regions (isospin angle $\theta_i =0$), blue is
the AB flux phase ($\theta=\half \pi$) and red the BA flux phase ($\theta=-\half\pi$). The
drawn $\cos(2\theta_{i_x})$ profile shows the boson density as a function of the
$x$-coordinate $i_x$ in the unit cell. Here,  $\theta_i$ is the isospin angle of the fermionic
mean field state $U_{ij}$. Note that the bosonic isospin angle as defined in Section
\ref{sect:psds}, Eq. \ref{MFstate}, is in this case equal to $\theta_{bos} = \theta_i -
\half\pi$. Hence, $\langle b_1 \rangle = \langle b_2 \rangle $ still holds in the
superconducting state. It is seen that the hole-rich region forms an antiphase boundary for
the SF-liquid state in between the superconducting stripes.}} \label{colourspiral}
\end{figure}

We propose a way in which the $SU(2)$ mean field theory can describe anti-phase boundaries,
within the spin liquid states descending from the Mott insulator. The first observation is
that for zero doping, the dSC or SFP state is just a gauge fix within the same projective
symmetry group. Let us now consider a gauge in which the spinon mean field $U_{ij}$ Eq.
(\ref{uij}) rotates over the whole isospin sphere,
\begin{equation}
U_{ij}=
\exp(i(-1)^I \theta_i \tau^1)\left (
      \begin{matrix}
     -\chi & \Delta  \\
       \Delta              & \chi
      \end{matrix}
 \right)\exp(-i(-1)^I \theta_i \tau^1)
\end{equation}
by a harmonically varying isospin angle
\begin{equation}
\theta_i = \mathbf{Q} \cdot i = q i_x, \mbox{\ \ \  ordering vector $\mathbf{Q}$ in $x$
direction}.
\end{equation}
In this way, the SFP state is smoothly connected to a dSC state. Observe that this state is in
the same PSG for zero doping. Then the idea is that for underdoped samples this spiral state
might be lower in energy than the phase separated state for the homogeneous $d+s$-wave
superconductor. A cartoon representation is given in the figure \ref{colourspiral}.

The peculiar feature of the isospin spiral is that in the $SU(2)$ gauge theory the
charge-density wave is made out of a superconductor, which is not the case in the large-$S$
limit. The antiferromagnetic domains are now replaced by a spin liquid, carrying nodal
fermions, a feature absent in the antiferromagnet. As the nodal fermions exist in both the dSC
and SF phases, a very promising perspective is opened up, supported by experiments.

The first support comes from the results from Fujimori and coworkers \cite{fujimori} and Z.X.
Shen and collaborators \cite{arpesreviewshen} for LaSCO. The chemical potential shift
measurements of Fujimori in underdoped LaSCO are compatible with the existence of charge
order, with Shen finding similar results. Furthermore, in the Nd-doped cuprates, clear
features of static stripes are measured already in the nineties \cite{tranquada95,zhou99}. On
the other hand, existence of nodal fermions in underdoped cuprates is found as well
\cite{arpesreviewshen}. The combination of these results seem to indicate  the coexistence of
striped charge order with nodal fermions. This idea is backed by recent results from  the
group of J.C. Davis \cite{seamus}, reporting that charge order and nodal fermions can coexist.

The $SU(2)$ gauge theory is able to capture 'Mottness', $d$-wave superconductivity and the
protection of nodal fermions. The framework of the isospin spiral state in $SU(2)$ mean field
theory forms an excellent explanation to explain the mystery why nodal fermions should exist
in a strongly correlated background. This is a promising motivation to study the stability of
the isospin spiral mean field states in underdoped cuprates.

\begin{figure}
\centering \rotatebox{0}{
\resizebox{8.3cm}{!}{%
\includegraphics*{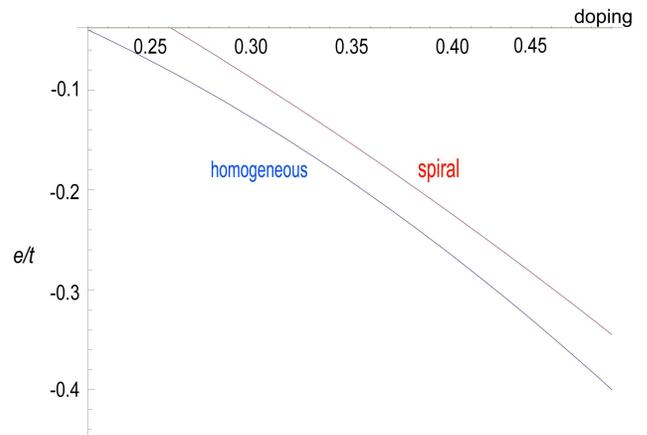}}}
\caption{\small{ The mean-field energy of the isospin spiral state as a function of doping.
The $Q=\frac{2\pi}{8}$ spiral state is taken, as depicted in Figure (\ref{colourspiral}), such
that the unit cell is $2\times 8$. For the $t-J$-model with $J=0.3t$, $t=0.44$ meV, the
homogeneous state is lower in energy, but not very much. Furthermore, one can see that the
isospin spiral state does not allow nonzero doping in the phase separation regime. This should
be viewed as an artefact of the $t-J$ model, however: any more realistic model, admitting
stripes, might make the isospin spiral state more favourable in energy in the phase separation
regime.}} \label{homvsinhom}
\end{figure}

As a first attempt, we adapted our program such that it can incorporate inhomogeneous
constraints, within the harmonic approximation. We used a 2$\times$8 unit cell, as suggested
in Figure \ref{colourspiral}. Via simulated annealing, a self-consistent solution can be found
for chemical potentials $\mu$ admitting inhomogeneous boson densities. In Figure
(\ref{homvsinhom}), the mean field energy as a function of average doping is plotted ,
together with the energy of the homogeneous solution.

It can be seen that the homogeneous states are better in energy in the homogeneous regime
($x>20\%$ for $J/t=0.3$). One expects that in the phase separation regime, the isospin spiral
state wins. Unfortunately, the spiral state is energetically a little bit less favourable the
homogeneous state, but this is an artefact of the unrealistic $t-J$-model. Any model which is
closer to the experimental reality, admitting stripes in the underdoped regime, are expected
to favour the spiral states. This hope is corroborated by the fact that the energy differences
with the homogeneous state are small, i.e., a couple of percent. The direction of future
research is clear: do models supporting stripes, also support isospin spiral states in the
$SU(2)$ slave boson formulation? This question is of high importance in the underdoped regime.
In that regime, it is interesting to see whether there can be made a ``Yamada-plot" of the
isospin speed $Q$ as a function of doping $x$, since $Q$ induces period $2Q$-stripes.

\begin{figure}
\centering \rotatebox{0}{
\resizebox{7.3cm}{!}{%
\includegraphics*{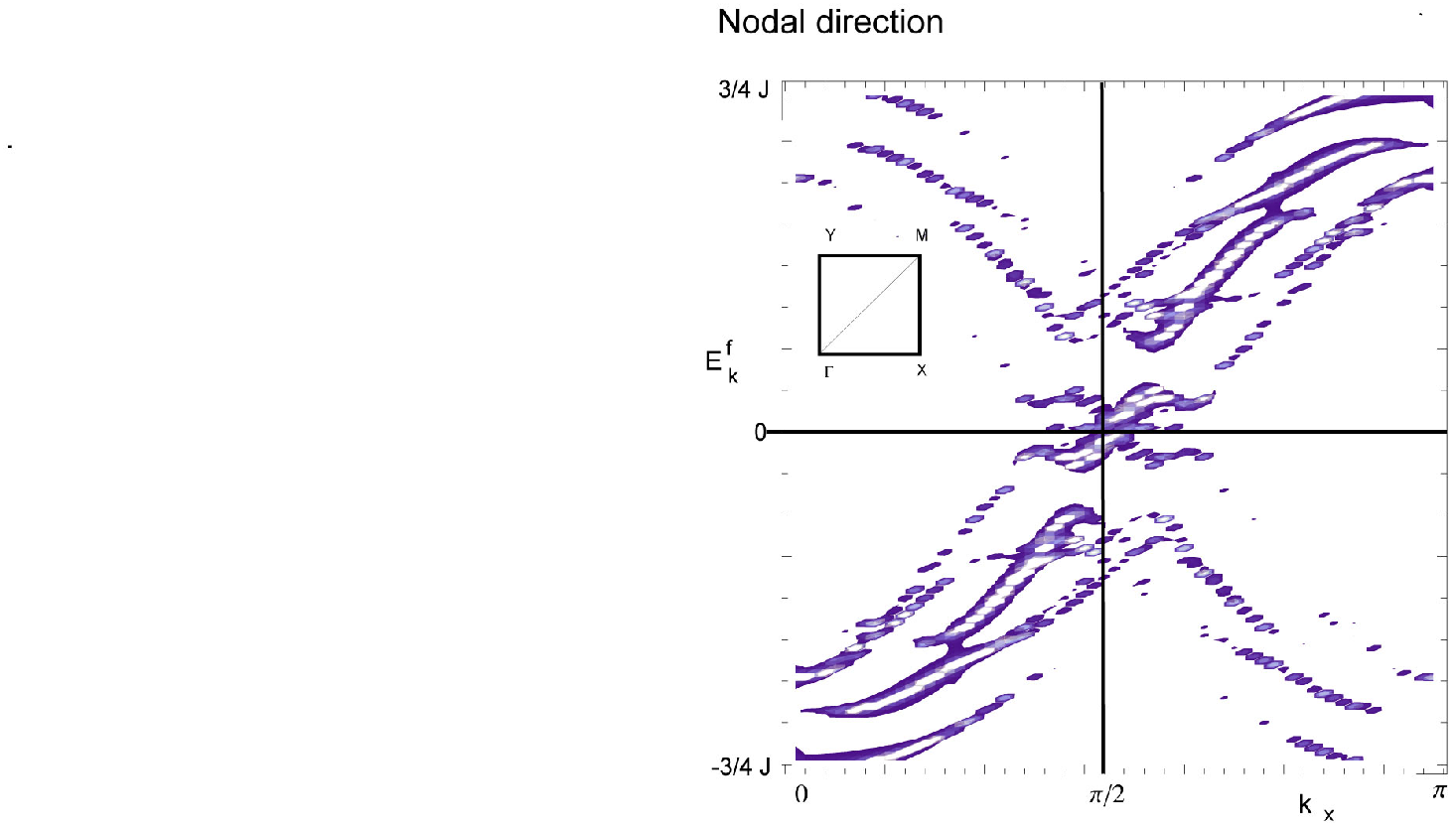}}}
\caption{\small{The dispersion of the spinons in the $Q=\frac{2\pi}{8}$ spiral state, on a
$8\times 2$ unit cell. The dispersion is along the nodal direction. The energy resolution is
$\Delta \hbar\omega/J = 1/160$, and the momentum resolution is $\Delta k /a = 1/128 $.  The
chemical potential was chosen such that the average doping was $26\%$, with variations
(stripes) in hole density of 2\%. The dispersion crosses the spinon Fermi surface at
$(\half\pi,\half\pi)$. It is seen that in contrast to usual belief, there is no appreciable
Umklapp scattering. This result supports the idea that by the isospin spiral state, the nodal
fermions coexist with stripes.}} \label{nodalfermions}
\end{figure}

In addition, we can shed some light on the issue of the coexistence of nodal fermions with
stripes. We determined numerically the dispersion relations for the spinons for the
$Q=\frac{2\pi}{8}$ spiral, on a $8\times 2$ unit cell. We only took the spinon part of the
propagator Eq.(\ref{convolprop}) into account. The result is shown in
Fig.(\ref{nodalfermions}).

The surprising result is that there is no (appreciable) Umklapp scattering in the spinon
sector, at this resolution at least. It is interesting to calculate the full electron spectral
density, taking both the inhomogeneities in the boson condensate and the incoherent phonon
part into account. We expect that the Umklapp might vanish altogether.

The Figure \ref{nodalfermions} might not be decisive, but very promising as to the idea that
nodal fermions and stripes are not as mutually exclusive as thought for a long time in the
high-$T_c$ community. Further effort should be put in more precisely mapping out the nodal
dispersion, especially in the underdoped regime.

In summary, we discussed how the constraint structure of the mean field $SU(2)$ gauge theory
of the doped Mott insulator forces the holons to have a hard core. This was demonstrated by
considering the empty limit. The empty limit was shown to produce the correct mean field
energy, using the hard-core condition, and the correct single-electron propagator. As a
surprise, we showed that an $s$-wave pairing in the spinon sector is inevitable, and grows
linearly with doping. Then the mean field behaviour for intermediate dopings was shown to
capture phase separation tendencies in the underdoped regime, being quantitatively in accord
with both numerical results and experiments. The single-particle propagator was shown to
display phonon modes.

We put forward indications that the $SU(2)$ gauge theory is able to capture both 'Mottness'
and $d$-wave superconductivity in a unifying picture, introducing the concept of the isospin
spiral. This idea might offer the possibility of the protection of nodal fermions, as the
calculated spinon sector of the single electron spectral function seems to indicate. The
framework of the isospin spiral state in $SU(2)$ mean field theory forms an excellent
explanation to explain the mystery why nodal fermions should exist in a strongly correlated
background. This is a promising motivation to study the stability of the isospin spiral mean
field states in underdoped cuprates further, together with making predictions for ARPES
measurements.

\textbf{ Acknowledgements:}   This work was financially supported by the Dutch Science
Foundation NWO/FOM.

\end{document}